\documentstyle[preprint,aps,floats]{revtex}
\tightenlines
\input psfig.tex
\begin{document}
\def\ba{\begin{eqnarray}}
\def\ea{\end{eqnarray}}
\def\be{\begin{equation}}
\def\ee{\end{equation}}
\def\({\left(}
\def\){\right)}
\def\[{\left[}
\def\]{\right]}
\def\lagrange {{\cal L}}
\def\del {\nabla}
\def\d {\partial}
\def\Tr{{\rm Tr}}
\def\half{{1\over 2}}
\def\fourth{{1\over 8}}
\def\bibi{\bibitem}
\def\S{{\cal S}}
\def\xx{\mbox{\boldmath $x$}}
\newcommand{\labeq}[1] {\label{eq:#1}}
\newcommand{\eqn}[1] {(\ref{eq:#1})}
\newcommand{\labfig}[1] {\label{fig:#1}}
\newcommand{\fig}[1] {\ref{fig:#1}}
\newcommand\bigdot[1] {\stackrel{\mbox{{\huge .}}}{#1}}
\newcommand\bigddot[1] {\stackrel{\mbox{{\huge ..}}}{#1}}
\title{The Scalar, 
Vector and Tensor Contributions to CMB anisotropies
from Cosmic Defects
} 
\author{
Neil Turok \thanks{email:N.G.Turok@damtp.cam.ac.uk},
Ue-Li Pen \thanks{email:upen@cfa.harvard.edu} and Uro\v s Seljak
\thanks{email:useljak@cfa.harvard.edu}} 
\address{${}^*$
DAMTP, Silver St,Cambridge, CB3 9EW, U.K.\\
 $^\dagger$Harvard College Observatory, 60 Garden St., Cambridge MA 02138\\
 ${}^\ddagger$ Harvard-Smithsonian Center for Astrophysics,
 60 Garden St., Cambridge MA 02138}
\date\today 
\maketitle

\begin{abstract}
Recent work has emphasised the importance of 
vector and tensor contributions to the large scale
microwave anisotropy fluctuations
produced by cosmic defects. In this paper we provide 
a general discussion of these contributions, 
and how their magnitude is constrained by the 
fundamental assumptions of causality, scaling, and 
statistical isotropy. We discuss an analytic  model which
illustrates and explains how the ratios of isotropic and anisotropic
scalar, vector and tensor microwave anisotropies are determined. 
This provides a check of the results from large scale numerical 
simulations, confirming the numerical finding that
vector and tensor modes provide 
substantial contributions to
the large angle anisotropies. This leads
to a suppression of the scalar normalisation and 
consequently of the Doppler peaks.
\end{abstract}
\vskip .2in

\section{Introduction}
The idea that the breakdown of 
some fundamental symmetry and the consequent field ordering 
might be responsible for structure formation in the 
universe is an attractive one. Recently we have performed 
the first 
complete calculations of the power spectra
of perturbations in symmetry breaking theories,
including global cosmic strings, monopoles 
and texture \cite{us1}, \cite{pap2}. These calculations revealed  that
vector and tensor modes give a larger contribution to the 
large scale anisotropies than previously suspected, and 
that their fractional contributions to the total microwave 
anisotropy power spectrum 
are comparable for each theory considered (see Figure 1).
Simultaneous work on local strings \cite{stebbins} has 
produced compatible 
conclusions.

The main implication of the large vector and tensor contribution
on large angular scales 
is in reducing the normalisation of the
scalar perturbations, which are responsible for the
Doppler peaks. Once the vector and 
tensor contributions are properly included, 
the height of the Doppler 
peaks are low relative to the large angular scale Sachs-Wolfe 
plateau\cite{us1}.

The present paper represents  an analytical attempt to explain why
vector and tensor contributions are substantial 
on large angular scales, using only
the most fundamental properties of
the simplest defect theories, namely scaling, causality and statistical
isotropy. We illustrate the arguments through comparison with
the results of 
numerical computations \cite{us1}.

\begin{figure}
\centerline{\psfig{file=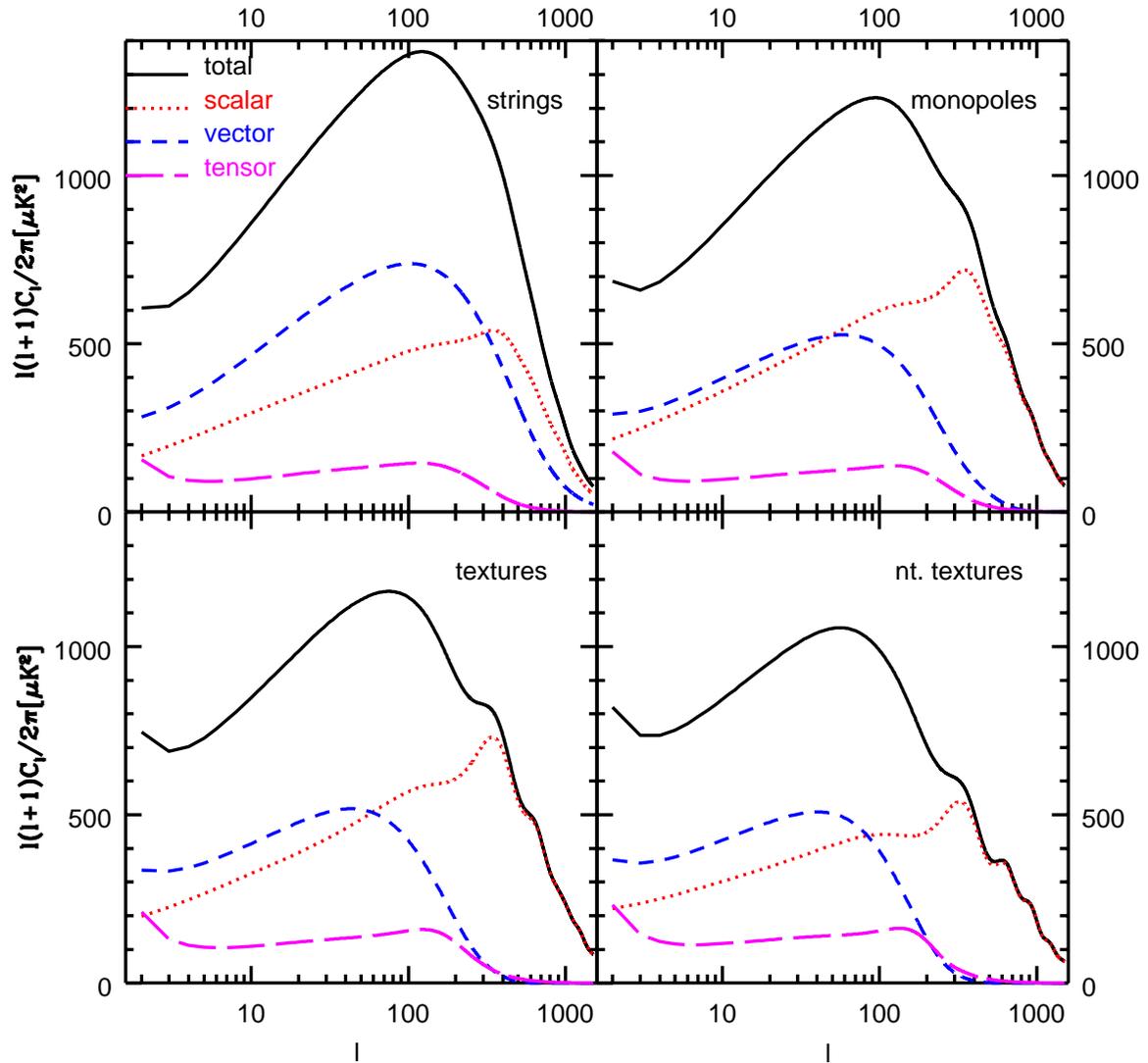,width=6.in}}
\caption{The contributions to the total anisotropy power spectrum from
scalar, vector and tensor components, in the theories
of global strings, monopoles, texture and nontopological texture
(taken from ref. [1]).}
\labfig{comp1}
\end{figure}

\section{Causality and Analyticity}

As discussed in \cite{pst}, all perturbation power spectra are 
determined by
the unequal time 
correlator (UETC) of the defect source stress energy tensor $\Theta_{\mu
\nu}$:
\newcommand{\cac}{{\cal C}}
\newcommand{\bk}{{\bf k}}
\be
\langle \Theta_{\mu \nu} (\bk,\tau) \Theta_{\rho \lambda}(-\bk,\tau')
\rangle \equiv {\cal C}_{\mu \nu,\rho \lambda} (k,\tau,\tau')
\labeq{uetc}
\ee
where $\tau$, $\tau'$ denote conformal time, and $k$ comoving wavenumber.
Note that (\ref{eq:uetc}) is real because complex conjugation 
is equivalent to the replacement $\bk \rightarrow -\bk$. The correlators
are invariant under this replacement because 
the statistical ensemble is rotation invariant.

Causality means that
the real space correlators of the 
fluctuating part of $ \Theta_{\mu \nu}$ 
must be zero for $r>\tau+\tau'$ \cite{ntcausal}.
Scaling dictates that in the pure matter or radiation eras
$\cac_{\mu \nu,\rho \lambda}
\propto \phi_0^4 /(\tau \tau')^{1\over 2} c_{\mu \nu,\rho \lambda}
(k\tau,k\tau')$, where $\phi_0$ is the symmetry breaking scale
and $c$ is a dimensionless scaling function. 
Finally, $ \Theta_{\mu \nu}$ must obey the equations for
stress energy conservation 
with respect to the background metric (see next Section).
 These
provide two linear constraints on the four scalar
components of the source. Any pair determines the other two 
up to possible integration constants. In the matter era
the pair
$\Theta$ and $\Theta^S$ \cite{pst} provides a
convenient choice, allowing
an analytical integral solution to the linearised
Einstein equations. But for work 
including the matter-radiation transition 
\cite{us1} the pair $\Theta_{00}$  and $\Theta^S$ 
is better,
because it results 
in the correct redshifting away of all components 
of the source stress energy 
inside the horizon. In this paper we shall use 
both pairs -
$\Theta$ and $\Theta^S$ in our analytical discussion 
of an `incoherent' model, and $\Theta_{00}$  and $\Theta^S$
for a numericaly solved `coherent' model. 
In the former case, we shall constrain 
$\Theta$ and $\Theta^S$  so that
on subhorizon scales the source is 
negligible (see Section IV).

The unequal time correlator in $k$ space is the Fourier 
transform of the real space correlator:
$\langle \Theta_{ij} (\bk,\tau) \Theta_{kl}(-\bk,\tau')
\rangle = 
\int d^3 {\bf r} e^{-i{\bf k.r}} \langle \Theta_{ij} (r,\tau) \Theta_
{kl}(0,\tau')\rangle$. 
The integral is
finite because the real space correlator has compact support, 
and it follows that
the unequal time correlators
are analytic in $\bk$ for all finite $k$. 
They may thus be expanded as Taylor
series in the Cartesian components $k_i$ about $k_i=0$. 
As $k_i$ tends to zero, 
isotropy and symmetry 
impose 
\be
{\rm Lim} \quad \bk \rightarrow 0 \quad 
\langle \Theta_{ij}(\bk,\tau) \Theta_{kl}(-\bk,\tau') 
\rangle =  A \delta_{ij} \delta_{kl} + B(\delta_{ik} \delta_{jl}
+\delta_{il} \delta_{jk})
\labeq{smallk}
\ee
with $A$ and $B$ independent of $\bk$.

The trace scalar, anisotropic scalar,
vector and tensor components of a tensor $T_{ij}$ are given by
\ba
T_{ij}(\bk) 
&=&{1\over 3}\delta_{ij} T + (\hat{k}_i \hat{k}_j - {1\over 3}
\delta_{ij}) T^S + (\hat{k}_i T^V_{ j} +\hat{k}_j T^V_{i}) + T^{T}_{  ij}\cr
T_i^V k_i &=& k_iT^T_{ij} = T^T_{ij} k_j = T^T_{jj} = 0,
\labeq{svtd}
\ea
where $\hat{k}_i \equiv k_i/k$.
Expressing the trace $T$, $T^S$, $T_i^V$ and $T_{ij}^T$ in terms of
$T_{ij}$ (see e.g. \cite{pst}) one finds that the only 
nonzero correlators consistent with 
statistical isotropy and homogeneity 
are $\langle TT \rangle$, $\langle TT^S \rangle$,
$\langle T^S T^S \rangle$, $\langle T^V_i T^V_j \rangle$ and
$\langle T^T_{ij} T^T_{kl} \rangle$. From
(\ref{eq:smallk}) one can compute  
the small $k$
power spectra of the anisotropic scalar,
vector and tensor stresses. One finds they are in the ratios
\be
\langle |\Theta^S|^2
\rangle :\langle |\Theta^V_i|^2
\rangle : \langle |\Theta^T_{ij} |^2
\rangle = 3:2:4
\labeq{ratios}
\ee
where all indices are summed.
Thus in a causal theory anisotropic scalar, vector and tensor
stresses have white noise components at small $k$ with 
related amplitudes.
A similar argument shows that the correlator 
$\langle \Theta_{00} \Theta_{ij} \rangle \sim {\rm C} \delta_{ij}$ at
small $k$, implying that 
$\langle \Theta_{00} \Theta^S \rangle $
vanishes like  $k^2$ at small $k$. Likewise 
$\langle \Theta \Theta^S \rangle $
vanishes like  $k^2$ at small $k$. So for either of the two
choices discussed above, 
the two  scalar source components are uncorrelated outside the horizon.

\begin{figure}
\centerline{\psfig{file=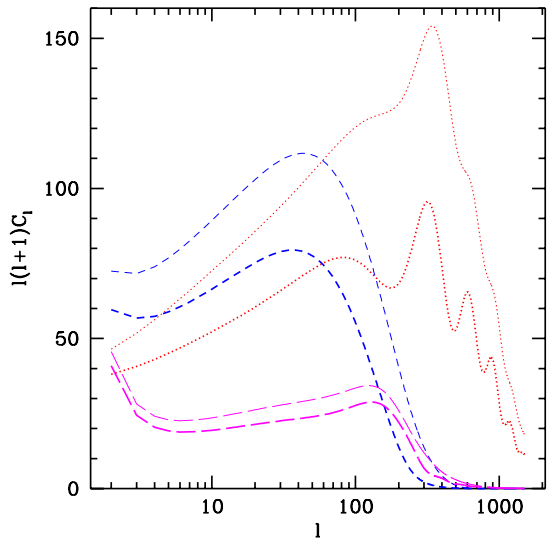,width=3.in}}
\caption{The importance of the long wavelength  modes in 
the anisotropy power spectra
from cosmic textures. The power spectra due to 
scalar (dotted), vector (dashed) and tensor (long-dashed) 
components of the sources 
are  compared to those where the source stress energy
components $\Theta_{00}$ and $\Theta^S$ as 
well as the vector and tensor stresses are set zero
for all $k\tau>5$. The 
upper curves show the full spectra, 
the lower ones the results where the cutoff is imposed.}
\labfig{comp4}
\end{figure}
\section{Superhorizon Modes}

In the cosmic defect theories, perturbations are
predominantly produced on the horizon scale. 
Studies show that the unequal time correlators 
take the predicted white noise form for $k\tau <5$ or
so, and decline strongly at larger $k\tau$. 
To the extent that the horizon scale modes
reflect the causality constraints discussed above,
the latter translate into definite 
relations between the scalar, vector and tensor perturbation 
power spectra. In Figures 2 and 3 we show the 
CMB anisotropy power spectra calculated in the 
cosmic global string and texture theories respectively, 
with and without a cutoff where we switch off
the source stress tensor for $k\tau >5$. The Figures show that
in the texture theory the effect of suppressing the 
source for $k\tau >5$ is relatively minor. For strings, there
is a larger effect, but even here 
the {\it ratios} of scalar to vector to tensor anisotropies are 
not much affected. We conclude that the
contributions from $k\tau <5$, which we shall term 
superhorizon modes, are certainly important in both theories and
give at the least a rough measure  of the 
importance of the scalar, vector and tensor contributions
to the  large angle anisotropies.

\begin{figure}
\centerline{\psfig{file=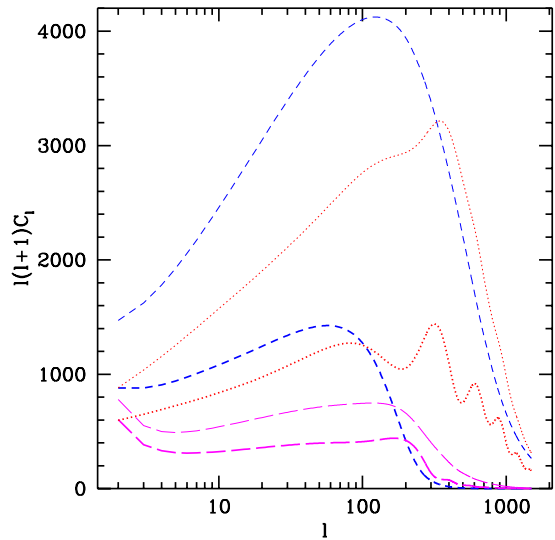,width=3.in}}
\caption{As in Figure 2 but for global strings.}
\labfig{comp5}
\end{figure}

\section{Integral Constraints}

The fact that a cutoff on subhorizon scales
does not greatly affect the large angle $C_l$ spectrum has important 
implications. It means that the short distance 
structure of the individual
defects is not important in determining the qualitative 
character of the 
large angle anisotropies, such as the relative scalar/vector/tensor
contributions. 

Consider the effect of modelling the sources using a `smoothed' 
$\Theta_{00}$ tensor, one where we impose a cutoff at $k\tau \sim 5$.
We feed in the smoothed $\Theta_{00}$
and $\Theta^S$ into the stress energy conervation equations:
\be
\dot{\Theta}_{00} + \frac{\dot{a}}{a}(\Theta_{00}+\Theta) = \Pi ;
\qquad
\dot{\Pi} + 2 \frac{\dot{a}}{a} \Pi = - \frac{k^2}{3} (\Theta + 2\Theta^S).
\label{eq:momc}
\ee
where $\Pi =\partial_i \Theta_{0i}$ and $\Theta$ and $\Theta^S$ are
defined in the previous Section. It is straightforward to see
that the solutions for $\Pi$ and $\Theta$ are well defined.

In this scheme, we deduce important relations between
$\Theta$ and $\Theta^S$. Equations
(\ref{eq:momc}) are easily integrated to obtain 
$\Theta_{00}$ in terms of $\Theta$ and $\Theta^S$:
exchanging the order
of the double integral we get
\be
\Theta_{00} = \tau^{-2} \bigl[ \int_0^\tau d \tau' 2 \tau'
\Theta 
+{1\over 3} k^2 \tau'^{4}({1\over \tau'}-{1\over \tau}) 
(\Theta + 2\Theta^S)
(\tau')\bigr],
\labeq{sole}
\ee
where we used $a(\tau) \propto \tau^2$ in the matter era.
But as argued, the smoothed 
$\Theta_{00}$ is identically zero inside
the horizon. 
It follows that
both the $\tau^{-2}$ and $\tau^{-3}$ coefficients
integrate to zero.
The former gives
\be
\int_0^\infty d \tau' \bigl[ 2 \tau' \Theta +{1\over 3} k^2 \tau'^3 
(\Theta + 2\Theta^S)(k,\tau')\bigr] = 0,
\label{eq:conb}
\ee
and the latter gives
\be
\int_0^\infty d \tau' \tau'^4 (\Theta + 2\Theta^S)(k,\tau') = 0.
\label{eq:cona}
\ee
This imposes a negative correlation between 
$\Theta$ and $\Theta^S$, and guarantees that $\Pi$ 
vanishes faster than $a^{-2}$  inside the horizon. 
The constraints (\ref{eq:conb}) and (\ref{eq:cona}) 
will turn out to be remarkably powerful 
when building models for the superhorizon 
components of $\Theta$ and $\Theta^S$.

\section{Perturbations in the Matter Era}

We wish to compute the large angular scale anisotropies 
produced in the matter era. For this purpose
we use the following integral solution to the linearised 
Einstein equations 
in a matter dominated universe \cite{pst}:
\ba
{\delta T \over T}({\bf n})|_{SW} &=&
-{1\over 2} \int_i^f d\tau h_{ij,0}(\tau, {\bf n}(\tau_0-\tau)) n^i n^j
;\qquad
h_{ij,0}=
h_{ij,0}^{\rm scalar} +
h_{ij,0}^{\rm vector} +
h_{ij,0}^{\rm tensor} \cr
h_{ij,0}^{\rm scalar} 
 &=&- 16 \pi G
\sum_{\bf k} e^{i {\bf k}.{\bf x}}
  \int_0^\tau d\tau' \bigl[
 {1\over 3}\delta_{ij} ({\tau' \over \tau})^6 (\Theta+ 2 \Theta^S)(\tau',
 {\bf k})
 - \hat{k}^i\hat{k}^j ({\tau' \over \tau})^4 \Theta^S(\tau', {\bf k})
 \bigr]\cr
h_{ij,0}^{\rm vector} &=& \sum_{\bf k} e^{i {\bf k}.{\bf x}}
(h_{i,0}^V \hat{k}_j + h_{j,0}^V \hat{k}_i) 
\qquad  h_{i,0}^V = 16 \pi G \int_0^\tau d\tau'
({\tau'\over \tau})^4 \Theta_i^V (\eta') \cr
h_{ij,0}^{\rm tensor} (\tau, {\bf x}) &=& 16 \pi G \int_0^\tau d\tau'
k^3 \tau'^4 \bigl[ G_1(\tau') \dot{G}_2(\tau)- G_2(\tau') \dot{G}_1(\tau)
\bigr]
\Theta_{ij}^T (\tau' {\bf x}) \cr
G_1(\tau) &=&
{{\rm cos} (k\tau) \over (k\tau )^2}
-{{\rm sin} (k\tau) \over (k\tau )^3}; \quad
G_2(\tau) =
{{\rm cos} (k\tau) \over (k\tau )^3}
+{{\rm sin} (k\tau) \over (k\tau )^2}; 
\labeq{exact}
\ea
where $G_1$ and $G_2$ are the two homogenous solutions to the tensor 
(gravity wave) equation.

The model we shall consider is one in which the components of 
$\Theta_{ij}$ have the following unequal time autocorrelators:
\ba
(16 \pi G)^2 \langle \Theta (\bk,\tau) \Theta(-\bk,\tau')
\rangle &=& \theta(\epsilon-k\tau) \delta(\tau-\tau') {\cal A}\cr
(16 \pi G)^2 \langle \Theta^{S,V,T} (\bk,\tau) \Theta(-\bk,\tau')^{S,V,T}
\rangle &=& \theta(\epsilon-k\tau) \delta(\tau-\tau') {\cal A}^{S,V,T}
\labeq{strcorr}
\ea
where $\theta$ is the Heaviside function, and we define 
$\langle \Theta^T (\tau) \Theta^T (\tau') \rangle \equiv
{1\over 4} \langle  \Theta_{ij}^T (\tau)  \Theta_{ij}^T (\tau') \rangle$,
and $\langle  \Theta^V (\tau)  \Theta^V (\tau') \rangle \equiv
{1\over 2} \langle  \Theta_{i}^V (\tau)  \Theta_{i}^V (\tau') \rangle$,
with all indices summed.

The sources are 
nonzero only on `superhorizon' scales ($k\tau < \epsilon$) and
they are uncorrelated except at equal times. 
This latter property means that the 
model is `totally incoherent', in the terminology of 
ref. \cite{albrecht}.
These correlators are not strictly 
causal - in real space they take the form 
$r^{-3} \bigl[{\rm sin}x  - x  
{\rm cos} x]$ where $x= \epsilon r/\tau$ - but they are 
small and 
oscillatory 
beyond  $r \sim \tau$ for
$\epsilon=5$. So the violations of causality are small.
Rotational invariance forbids any cross-correlation between
scalar, vector or tensor modes. There is however one more 
allowed cross correlator, namely that between the 
isotropic and anisotropic stresses. 
The argument given in Section III implies that
$\langle \Theta\Theta^S \rangle $ vanishes as $k^2$ for small $k$,
but it cannot be zero because of the constraint (\ref{eq:cona}).
We choose to model it as
\be
(16 \pi G)^2 \langle \Theta (\bk,\tau) \Theta^S(-\bk,\tau')
\rangle = \theta(\epsilon-k\tau) \delta(\tau-\tau') 
\bigl({k\tau\over 
\epsilon}\bigr)^2
{\cal A}^{\Theta S}
\labeq{ccorr}
\ee
If we now compute the equal time correlator of the 
the constraint (\ref{eq:cona}), we determine
\be
{\cal A}^{\Theta S} = -{11\over 36} ({\cal A}+4{\cal A}^S ).
\labeq{ccorrv}
\ee
Similarly we compute the equal time correlator of
(\ref{eq:conb}) and 
obtain 
\be
({4\over 3} + {4\over 15}\epsilon^2 + {1\over 63}\epsilon^4){\cal A}
+ ({8\over 21} \epsilon^2 + {4\over 81} \epsilon^4) {\cal A}^{\Theta S} 
+{4\over 63} \epsilon^4 {\cal A}^{S}=0.
\labeq{ccorrva}
\ee
These equations
yield ${\cal A}^{\Theta S}=-2.47{\cal A}^S$
and ${\cal A}= 4.07 {\cal A}^S$ for $\epsilon=5$. 
At $k\tau=5$, there is
a mild inconsistency with the bound ${\cal A}^{\Theta S}<\sqrt{{\cal A} 
{\cal A}^S}$, so we shall adopt ${\cal A}^{\Theta S}=-2{\cal A}^S$ 
and ${\cal A}= 4{\cal A}^S$.

\section{The Delta Function  Approximation} 

The procedure is simple in principle: the correlators 
(\ref{eq:strcorr}) translate into correlators of the 
metric perturbations and thus into correlators of the 
temperature perturbations, equivalent to the anisotropy 
power spectrum $C_l$. But in order to compute the relevant integrals 
analytically, 
we shall make two approximations. 
The first is that we shall replace 
metric perturbation unequal time correlators under
$\tau$ integrals 
using
the following formula:
\ba
\langle \dot{A}(\tau) \dot{B}(\tau') 
\rangle &\rightarrow& \delta(\tau-\tau') {d \over d \tau} \langle A(\tau) B(\tau) 
\rangle 
\labeq{approx}
\ea
The weighting function is chosen so that 
the integrals $\int_0^{\tau_f }
d\tau 
\int_0^{\tau_f} d \tau'$ of both sides are guaranteed to be equal 
for all $\tau_f$. 
The formula is also invariant under changing 
variables from $\tau$ to any other function $f(\tau)$.
The second approximation is to use the fact that 
the Greens functions in (\ref{eq:exact})
fall off strongly with $\tau$. This means that 
the metric perturbations fall off 
rapidly beyond $k \tau = \epsilon$, which 
justifies us simply setting them 
zero beyond that point.

\section{CMB Anisotropies}

In the usual way
we expand the microwave sky temperature in spherical
harmonics $\delta T/T = \sum a_{lm} Y_{lm}(\theta,\phi)$, 
and compute $C_l = \langle |a_{lm}|^2 \rangle$.
The  formulae for the contribution to the integrated 
Sachs Wolfe effect 
from trace scalar and anisotropic scalar contributions are
\cite{crit}:
\be
C_l^{scalar} = {1 \over 2 \pi}
 \int_0^\infty k^2 dk
\langle
\bigl[\int_0^{\tau_0}
d\tau  \bigl(  {1\over 3} \dot{h}_1(\tau) + \dot{h}_2(\tau)
{d^2 \qquad \over  d (k \Delta \tau)^2} 
\bigr) j_l(k\Delta \tau) \bigr]^2 
\rangle
\labeq{scalcleq}
\ee
where $\Delta \tau = \tau_0-\tau$, $\tau_0$ is the conformal time
today and as above, $\langle \rangle$ denotes ensemble averaging.
The scalar metric perturbation components 
are given from (\ref{eq:exact}):
\ba
\dot{h}_{1} &=&  -16 \pi G \int d \tau' (\tau'/\tau)^6 
(\Theta+2\Theta^S)(\tau') \cr
\dot{h}_{2} &=&  -16 \pi G \int d \tau' (\tau'/\tau)^4 \Theta^S(\tau') 
\labeq{hdotsol}
\ea
with ${\bf k}$ dependence implicit.

The vector and tensor contributions to $C_l$ are \cite{crit}
\ba
C_l^V &=& 
 {2  \over \pi} \int_0^\infty k^2 dk l(l+1) 
\langle
\bigl[\int_0^{\tau_0}
d\tau  \dot{h}^V(\tau)
{d\qquad \over  d (k \Delta \tau)} \bigl(j_l(k\Delta \tau)/ k\Delta \tau \bigr)\bigr]^2 
\rangle \cr
C_l^T &=& {1 \over 2 \pi}
 \int_0^\infty k^2 dk {(l+2)! \over (l-2)!} 
\langle
\bigl[\int_0^{\tau_0}
{d\tau \over k^2 \Delta \tau^2}  \dot{h}^T(\tau) j_l(k\Delta \tau) \bigr]^2 
\rangle 
\labeq{cleq}
\ea
where 
$\langle \dot h^T (\tau) \dot h^T (\tau') \rangle \equiv 
{1\over 4} \langle \dot h_{ij}^T (\tau) \dot h_{ij}^T (\tau') \rangle$, 
and $\langle \dot h^V (\tau) \dot h^V (\tau') \rangle \equiv 
{1\over 2} \langle \dot h_{i}^V (\tau) \dot h_{i}^V (\tau') \rangle$,
with all indices summed. 

We now compute the relevant metric perturbation correlators:
from (\ref{eq:strcorr}),  (\ref{eq:ccorrv})
and (\ref{eq:hdotsol}),
using ${\cal A}^{\Theta S}=-2{\cal A}^S$
and ${\cal A}= 4{\cal A}^S$ as discussed above, we
obtain 
\ba
\langle h_1(\tau)^2 \rangle &\approx& 
{1\over 40}  \tau^3 {\cal A}^S
\cr 
\langle h_2(\tau)^2 \rangle &=& {1\over 81} \tau^3 {\cal A}^S
\cr 
\langle h_1(\tau)h_2(\tau)\rangle 
&\approx& {1\over 217} \tau^3 {\cal A}^S \cr
\langle h^V(\tau)^2 \rangle &=& {1\over 81} \tau^3 {\cal A}^V \cr
\langle h^T(\tau)^2 \rangle &=& 
\int_0^\tau d \tau'
 G(\tau,\tau')^2 {\cal A}^T\cr
G(\tau,\tau') &=&
 k^3 \tau'^4 \bigl[ G_1(\tau') G_2(\tau)- G_2(\tau') G_1(\tau)
\bigr]
\labeq{happ}
\ea
where we have evaluated the scalar correlators at
$k\tau=5$, and the tensor modes 
$G_1$ and $G_2$ are given in equation  (\ref{eq:exact}). 
The tensor integral is straightforwardly performed, yielding 
\ba
 \langle h^T(\tau)^2 \rangle &=& {{\cal A}^T \over 60 k^3 z^6}
\bigl( {105\over 2} (1-z^2) {\rm sin}(2z)- 105z{\rm cos}(2z) +6 z^7
-14z^5-35z^3\bigr) 
\ea
where $z=k\tau$. This function is $\sim {\cal A}^T\tau^3/81$ at small
$k\tau$, identical to the vector expression. 
But for larger $k\tau$ it is suppressed, with the
suppression factor being 
$\approx 0.29$ at $k \tau =5$. 
The suppression 
is due to the oscillatory nature of the tensor compared to the vector
response.

We now compute the integrals in (\ref{eq:cleq}), starting with 
the tensor contribution $C_l^T$. The delta function
allows one of the $\tau$ integrals to be performed. Then we
change variables from $k$ to $x=k\Delta \tau$. The
Heaviside function 
gives the upper limit $x < \epsilon (\tau_0/\tau -1)$, or  $\tau <
\tau_0/(1+x/\epsilon)$. Exchanging orders of the 
integrals we find
\ba
C_l^T &=& {1 \over 2 \pi}
{(l+2)! \over (l-2)!} \int_0^\infty {dx \over x^2} j_l^2(x)
 \int _0^{\tau_0/(1+x/\epsilon)} {d \tau \over (\tau_0 -\tau)^3 }
{d \over d \tau}
\langle h^T(\tau)^2 \rangle 
\labeq{cltens}
\ea
For large $l$  the integral is dominated by large $x$, since 
$j_l(x) \sim x^l$ at small $x$.  But at large $x$, $\tau <<\tau_0$ and
the $\tau$ integral is trivial. 
Thus one finds at large $l$
\be
C_l^T \sim .29 {\epsilon^3 \over 81} {\cal A}^T  {l^4 \over 2 \pi}
\int_0^\infty {dx \over x^5} j_l^2(x) 
\sim 0.29 {\epsilon^3 \over  \pi} 
{\cal A}^T {2 \over 1215 l^2 } +o(l^{-3}).
\labeq{cltensf}
\ee
where we have used
\be
\int dx j_l^2(x) x^{n} = {\pi \over 2^{2-n}} {\Gamma(1-n) 
\Gamma(l+{1\over 2} +{n\over 2})
 \over \Gamma(1-{n\over 2})^2 \Gamma(l+{3\over 2
} -{n\over 2}) }.
\labeq{bessint}
\ee
We have computed the integral for the tensor contribution (\ref{eq:cltens})
at low $l$ 
using Mathematica, to check that the model reproduces the shape of
$l(l+1) C_l$ seen in the plots of Figure 1.
Figure 4 confirms that this
is indeed the case.

\begin{figure}
\centerline{\psfig{file=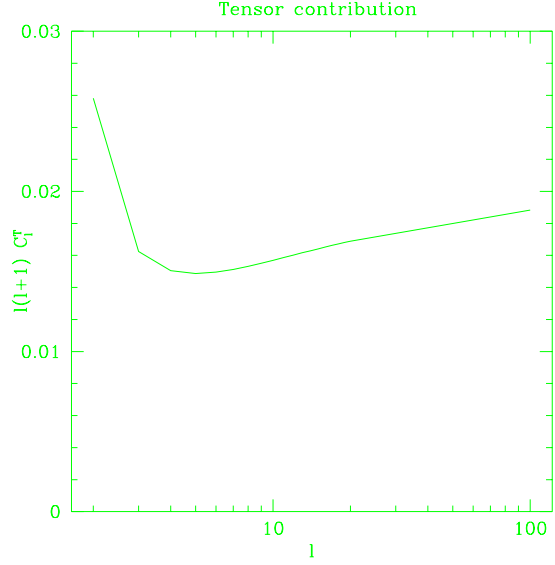,width=3.in}}
\caption{Tensor anisotropy power spectrum as computed in 
the analytical model presented here.}
\labfig{comp6}
\end{figure}

The vector integral is performed similarly, to obtain 
\be
C_l^V = {1\over 27} {\cal A}^V {2 l (l+1) \over \pi}
\int_0^\infty dx x^2 [(j_l(x)/x)']^2 
\bigl[
{\epsilon^2 \over 2 x^2} -{\epsilon\over x} +{\rm ln} (1+\epsilon/x)\bigr].
\labeq{clvec}
\ee
where prime denotes differentiation with respect to $x$. 
Approximating the square bracket with its leading large $x$ behaviour,
integrating by parts and using Bessel's 
equation for $j_l(x)$, we obtain at large $l$ 
\be
C_l^V 
\sim 
{\cal A}^V {2 \epsilon^3 \over 1215 \pi l^2 } +o(l^{-3}).
\labeq{clvectf}
\ee
The scalar contribution is evaluated using 
(\ref{eq:approx}) and (\ref{eq:hdotsol}), again making the 
large $x$ approximation, giving
\be
C_l^{scalar} = {\cal A}^S {1 \over 2 \pi}
\int_0^\infty x^2 dx ({1\over 120} j_l^2 +{1\over 81}
j_l''^2 +{4\over 217} j_l j_l'')(x) {\epsilon^3\over 2 x^3}. 
\labeq{claniso}
\ee
After integrations by parts and using Bessel's equations
we get
\be
C_l^{scalar} \approx {1\over 1248}  
{\cal A}^S {\epsilon^3 \over \pi l^2 } +o(l^{-3}).
\labeq{clsaniso}
\ee
Now the amplitudes ${\cal A}^S$, ${\cal A}^V$ and ${\cal A}^T$
are related via (\ref{eq:ratios}) in the ratio 
3: 1: 1. Thus 
in the various 
approximations we have made, the ratio of the 
scalar to
vector to tensor contributions to the large angle
anisotropies are 
\be
C_l^{scalar}:\quad C_l^{V}:\quad C_l^{T} = 1.46 :\quad  1: \quad 0.29
\labeq{res}
\ee
which is our main result.
The calculation demonstrates the relative importance of the 
vector and tensor modes, consistent with the
numerical results shown in Figure 1. 
Given the crude
nature of the model used, the agreement is actually 
surprisingly good. The weakest point in the model
is that it involves a free parameter $\epsilon$, 
and the $C_l$'s obtained are
proportional to $\epsilon^3$. It seems plausible that
$\epsilon$ should be the same for the scalar, vector and
tensor stresses, but we have not found any argument as
to why this should necessarily be true.

Let us summarise  the approximations and assumptions
implicit in (\ref{eq:res}):

\noindent
$\bullet$ We assumed
that superhorizon modes with $k\tau <5$ dominate.

\noindent
$\bullet$ We modelled the unequal time correlators as
delta functions with a horizon scale cutoff.

\noindent
$\bullet$ We 
made the approximation of pure matter domination for the
background spacetime. In this approximation the $C_l$ spectra 
obtained are scale invariant at large $l$.

\noindent
$\bullet$ We
replaced certain functions with amplitudes times delta functions
in order to perform the relevant integrals.


With all of these caveats, we feel that the model provides 
useful insight into the relative importance
of scalar, vector and tensor contributions to the 
large angle anisotropies. The model explains why 
vector perturbations dominate over tensors, and why 
the combined vector and tensor 
contribution is comparable 
to that from scalars.

\section{Numerical Solution of a Coherent Model}

As a further model we have considered the case of 
a completely coherent source in which the unequal
time correlators of $\Theta_{00}$, $\Theta^S$, 
 $\Theta^T$, and  $\Theta^V$ are all proportional to
the product of Heaviside functions, for example setting
\be
\langle \Theta_{00} (\bk,\tau) \Theta_{00}(-\bk,\tau')
\rangle = (\tau \tau')^{-{1\over 2}} 
\Theta(\epsilon - k\tau) \Theta(\epsilon - k\tau'). 
\labeq{nummodel}
\ee
Note that we need to model the white noise $\Theta^S$ 
contribution, but as mentioned above, the cross
correlator must vanish at small $k$. So in this model
we will assume that $<\Theta_{00} \Theta^S>$ is 
identically zero, and therefore solve for 
the $\Theta_{00}$ and  $\Theta^S$  contributions 
separately. Note that with this choice of variables, 
the constraints (\ref{eq:conb}) and (\ref{eq:cona}) are 
automatically satisfied.

\begin{figure}
\centerline{\psfig{file=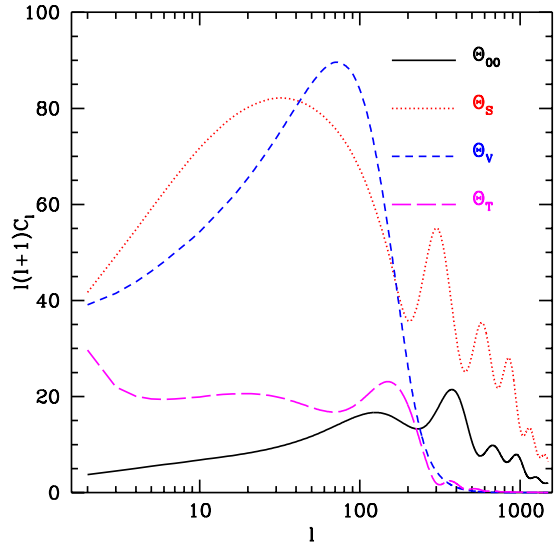,width=3.in}}
\caption{Angular power spectrum of anisotropies generated by  
a simple coherent model of scaling sources, correctly incorporating
the superhorizon constraints on the relative importance of
the anisotropic stress $\Theta^S$, vector $\Theta^V$ and 
tensor $\Theta^T$ perturbations to the source stress tensor. 
}
\labfig{model}
\end{figure}

We have used this model in the full Boltzmann code
developed in \cite{us1} as usual with
$\epsilon= 5$. 
The results are shown in Figure 5. The anisotropic
scalar, vector and tensor $C_l$'s have been scaled so that 
the stress tensor white noise superhorizon amplitudes
are in the correct ratios (\ref{eq:ratios}). 
$C_l^{00}$ has been scaled with $C_l^{S}$. It is a remarkably small
contribution. This may be understood by solving the 
equations for stress energy conservation for $\Theta$, 
whence one finds that $\Theta $ is actually very small
in the coherent model at horizon crossing $k\tau \sim 5$, so that 
from (\ref{eq:exact}) its contribution to the 
anisotropy is small.

The coherent model provides a useful comparison to the
previous 
incoherent model. The broad agreement
between the two models suggests that our main result 
(\ref{eq:res}) is actually insensitive to the 
detailed nature of the source.
An advantage of the coherent model is that
we can more easily incorporate
the matter-radiation transition, giving rise to 
departures from scale invariance in the $C_l$ spectrum 
qualitatively similar in character to those observed in 
the realistic source calculations.
And as seen in Figure 5 the model gives 
a reasonable  impression of the main features of the
realistic calculations in Figure 1, at least on large angular
scales.

\section{Conclusions}

In this paper we have developed a set of physically reasonable
 models for
the perturbations generated on superhorizon scales by causal 
sources. 
We gave some rigorous and some approximate arguments
that the large angular scale anisotropies due to the
scalar and vector plus tensor modes are in general
similar in magnitude.
If the vector and tensor contributions to 
the large angle anisotropies are large, the scalar 
normalisation is lower and 
the Doppler peaks due to scalar perturbations
are small compared to the large angle 
Sachs-Wolfe plateau.

Let us close by mentioning some
loopholes in the above arguments, which make it possible 
to circumvent the
conclusion that causal sources are unlikely
to have large Doppler peaks.

$\bullet$ If subhorizon modes with $k\tau >5$ dominate the
anisotropies then our arguments do not apply. 
Sources in this category have been explored by Durrer and
Sakellariadou \cite{durrer}.

$\bullet$ If the shape of the unequal time correlators, 
parametrised in our models by $\epsilon$ is different for
the scalar, vector and tensor components, then 
the $C_l$ contributions could be strongly affected,
since $C_l \propto \epsilon^3$.  One could imagine a model where 
$\epsilon$ for scalars was larger than for vectors and tensors, 
but even here one would probably not find sharp Doppler 
peaks, since 
increasing $\epsilon$ is likely to increase the 
incoherence of the source and thus smoothe out the 
Doppler peaks
(this is apparent in Figures 2 and 3). 

$\bullet$ One could consider sources like those in \cite{ntcausal}
in which the anisotropic stresses are by construction 
zero outside the horizon.
In such a model the superhorizon 
constraint (\ref{eq:ratios}) is satisfied
with all terms being zero. In the model of \cite{ntcausal}
this is true because the real space stress energy master 
functions
were taken to be {\it spherically symmetric},
clearly  a  special case.

$\bullet$ We have assumed perfect
scaling of the sources, and matter domination.
There is some
violation of scaling due to the matter-radiation 
transition, but this is a small effect on large angular scales. 
Stronger departures from scaling would
result from 
a non-minimal-coupling mass term $R\theta^2$ for the Goldstone bosons.
We are currently exploring this possibility.

\section{Acknowledgements}
We thank A. Albrecht, R. Battye and J. Robinson for helpful
discussions.

\end{document}